\documentclass[aps,prb,twocolumn]{revtex4}

\usepackage{amsmath}
\usepackage{amssymb}
\usepackage{natbib}
\usepackage{graphicx}
\usepackage{bm}

\newcommand{\m}{\textbf{m}}

\def\be{\begin{equation}}
\def\ee{\end{equation}}
\def\ba{\begin{eqnarray}}
\def\ea{\end{eqnarray}}
\def\rr{{\bf r}}

\def\w{\omega}

\def\e{\epsilon}

\def\p{\partial}
\def\jj{\textbf{j}}
\def\pp{\textbf{p}}
\def\rr{\textbf{r}}
\def\qq{\textbf{q}}
\def\vv{\textbf{v}}

\def\EE{\textbf{E}}
\def\BB{\textbf{B}}

\def\bra{\langle}
\def\ket{\rangle}

\begin{document}

\title{Chiral magnetic effect and natural optical activity in (Weyl) metals}
\date{\today}
\author{Jing Ma, D. A. Pesin}
\affiliation{Department of Physics and Astronomy, University of Utah, Salt Lake City, UT 84112 USA}
\begin{abstract}
We consider the phenomenon of natural optical activity, and related chiral magnetic effect in metals with low carrier concentration. To reveal the correspondence between the two phenomena, we compute the optical conductivity of a noncentrosymmetric metal to linear order in the wave vector of the light wave, specializing to the low-frequency regime. We show that it is the orbital magnetic moment of  quasiparticles that is responsible for the natural optical activity, and thus the chiral magnetic effect. While for purely static magnetic fields the chiral magnetic effect is known to have a topological origin and to be related to the presence of Berry curvature monopoles (Weyl points) in the band structure, we show that the existence of Berry monopoles is not required for the dynamic chiral magnetic effect to appear; the latter is thus not unique to Weyl metals. The magnitude of the dynamic chiral magnetic effect in a material is related to the trace of its gyrotropic tensor. We discuss the conditions under which this trace is non-zero; in noncentrosymmetric Weyl metals it is found to be proportional to the energy-space dipole moment of Berry curvature monopoles. The calculations are done within both the semiclassical kinetic equation, and Kubo linear response formalisms, with coincident results.
\end{abstract}
\maketitle

\section{Introduction}

The phenomenon of natural optical activity  was discovered by Arago in 1811, while observing passage of polarized light through crystals of quartz. This was the first example of a general family of phenomena related to optical activity, which is characterized by a material's different response to right- and left-handedly polarized light. Since then natural optical activity has become one of the most studied optical phenomena in molecules and crystals, with a wide spectrum of biomedical applications, ranging from determining sugar concentration in biological fluids, to studies of RNA and DNA molecules.

From modern perspective, natural optical activity in time-reversal invariant systems appears due to a linear in wave vector of the light wave spatial dispersion of the conductivity (or, equivalently, dielectric) tensor $\sigma_{ab}(\w,\qq)$:\cite{LL8}
\begin{equation}\label{eq:conductivity}
  \sigma_{ab}(\w,\qq)=\sigma_{ab}(\w)+\lambda_{abc}(\w)q_c.
\end{equation}
Since the third-rank tensor $\lambda_{abc}$ changes sign under inversion, the latter cannot be a symmetry of a material showing natural optical activity. The lack of an inversion center is a necessary, but not sufficient condition for natural optical activity.\cite{Nye} Noncentrosymmetric (lacking an inversion center) optically active materials are called \emph{gyrotropic}.

This work considers natural optical activity in metallic systems, since mechanisms of natural optical activity/gyrotropy in molecules and insulating crystals are well understood.\cite{BarronBook,KizelReview}   Our motivation comes from the observation that the so-called ``chiral magnetic effect'' (CME)~\cite{Vilenkin,Nielsen1983,Cheianov1998,Kharzeev2008,Son2013}, in particular discussed\cite{ZyuzinWu2012,Zyuzin2012,Goswami2013,Niu2013,Franz2013,ChenBurkov2013} in the context of Weyl (semi)metals,\cite{WanVishwanath2011,BurkovBalents2011,Turner2013,Hosur2013} is nothing but a particular case of optical activity.

Indeed, CME is defined as the existence of a current, $\jj$, flowing in response to a magnetic field, $\BB$, along the latter; in Fourier components, such a relation is written as
\begin{equation}\label{eq:jtoB}
\jj(\w,\qq)=\eta(\w,\qq)\BB(\w,\qq).
\end{equation}
Since this response exists only at finite frequencies in an equilibrium crystal,\cite{Niu2013,Franz2013,ChenBurkov2013} the Faraday's law $\BB=\qq\times \EE/\w$ ($c=1$ throughout this paper) allows to rewrite this relation as a particular case of spatial dispersion part of Eq.~(\ref{eq:conductivity}):
\begin{equation}\label{eq:lambdaideal}
  \lambda^{metal}_{abc}(\w)=-\frac{\eta(\w,\qq)}{\w}\e_{abc},
\end{equation}
where $\e_{abc}$ is the fully antisymmetric Levi-Civita tensor. At small frequencies, (the precise condition is to be discussed below) $\eta(\w,\qq)\approx \text{const}$, and thus dependence on frequency of $\lambda^{metal}_{abc}(\w)\propto1/\w$ is markedly different from $\lambda^{insulator}_{abc}(\w)\propto\w$ in the insulating case. The latter fact is expected, since in an insulator $\lambda_{abc}(\w)$ should be an analytic function of $\w$, and can be established on general grounds within band theory.~\cite{Natori1975,Malashevich} At high frequencies, the metallic and insulating cases cannot be distinguished by the frequency dependencies of respective $\lambda_{abc}$, hence we limit ourselves to the low-frequency regime in what follows.

In this work we extend the understanding of natural optical activity in metals in the following ways: $i$) We obtain an expression for tensor $\lambda_{abc}$ that determines the spatial dispersion of conductivity, Eq.~(\ref{eq:conductivity}), which is valid for arbitrary band structures at low frequencies, both from semiclassical kinetic equation, and from Kubo formalism. In general, we find that there are many conflicting results regarding this tensor in the literature. In particular, our results differ from those of Refs.~\onlinecite{OrensteinMoore2013,Goswami2014,HosurQi2015,Moore2015}; $ii$) Using the obtained results, we show that as far as Weyl metals are concerned, only the static limit of CME has a topological origin and a universal magnitude, which universally vanishes in an equilibrium crystal. In the dynamic limit, CME, being a particular part of current due to natural optical activity of the material, does exist, but is not universal: the tensorial structure of $\lambda_{abc}$ is more complicated than that of Eq.~(\ref{eq:lambdaideal}), and the magnitude of dynamic CME depends on the peculiarities of band structure (like its curvature). For the ideal case of a Weyl metal represented by a collection of particle-hole symmetric Weyl points, we show in full generality that the magnitude of the CME is determined by the energy-space dipole moment of Berry monopoles that correspond to these Weyl points; $iii$) Finally, we show that the possibility of CME-like response is not restricted to Weyl semimetals: In fact, the existence of Berry monopoles, and thus chiral anomaly, is not necessary for non-zero dynamic CME at all.

The rest of the paper is organized as follows: In Section~\ref{sec:semiclassics} we obtain a description of natural optical activity in metals based on the semiclassical kinetic equation. This Section also contains a discussion of the relation between the results of this paper, and other works. In Section~\ref{sec:examples} we consider several specific examples of model Hamiltonians that illustrate the results we obtained. In Section~\ref{sec:conclusion} we summarize our main results. Finally, in the Appendices we discuss the static CME within the semiclassical kinetic equation formalism, and present a derivation of the results from Section~\ref{sec:semiclassics} based on the Kubo formalism.

\section{Semiclassical theory of natural optical activity in metals}\label{sec:semiclassics}

In this section we describe the response of a time-reversal invariant metal to an electromagnetic field that varies slowly in  time and space, at linear response level. The main results of this Section are Eq.~(\ref{eq:gyrocurrent}) for the gyrotropic current, Eqs.~(\ref{eq:lambda}) and (\ref{eq:lambdaglobal}) for tensor $\lambda_{abc}$ that determines the spatial dispersion of conductivity, and Eqs.~(\ref{eq:gyrotensor}) and (\ref{eq:gyrotrace}) for the gyrotropic tensor and its trace.

 The formalism is borrowed from the Berry phase theory of semiclasiclassical transport~\cite{XiaoNiu}. Since we are interested in the leading effects of spatial dispersion of conductivity, we concentrate on current contributions that are proportional to the magnetic field, $B$, or linear-in-wave-vector contributions proportional to the electric field, $q E/\omega$. The Faraday's law dictates that these terms are of the same order of smallness in $q/\omega$ ratio for $q\to0$. We also disregard terms with higher order derivatives, since those are in general present in insulators with space groups that allow natural optical activity, e.g. tellurium,\cite{Nomura1960} while the terms we do consider only appear in systems with Fermi surfaces. We also focus on orbital mechanisms of natural optical activity, neglecting the spin contribution.\cite{noteonspin}

In order to calculate the current response to electromagnetic fields in the semiclassical regime, we need to express the electric current through the distribution function of electrons, and formulate a kinetic equation for the latter. We thus briefly recall known facts about semiclassical theory of electrons in crystals.

In what follows, we will denote the $p$-space Hamiltonian, the energy spectrum of the crystal, and the periodic parts of the corresponding Bloch wave functions in the absence of external fields with $h_\pp$, $\e_{n\pp}$, and $|u_{n\pp}\ket$, respectively. The gradient in the quasimomentum $\pp$-space is denoted with $\p_\pp$, and the derivative with respect to the $a^{\text{th}}$ component of $\pp$ with $\p_a$. The operator of gradient in real space is denoted with $\p_\rr$. For brevity, we set $\hbar=c=1$ throughout the paper.

At the semiclassical level, the dependence of periodic parts of Bloch functions on quasimomentum leads to the appearance of two objects,\cite{XiaoNiu} central for our discussion: The first is the Berry curvature of band $n$, ${\bf \Omega}_{n\pp}$,
\begin{equation}\label{eq:Omega}
{\bf \Omega}_{n\pp}=i\bra\p_\pp u_{n\pp}|\times|\p_\pp u_{n\pp}\ket;
\end{equation}
the other one is the orbital magnetic moment of quasiparticles, $\m_{n\pp}$,
\begin{equation}\label{eq:magneticmoment}
\m_{n\pp}=\frac {ie}2 \bra \p_\pp u_{n\pp}|\times (h_\pp-\e_{n\pp})|\p_\pp u_{n\pp}\ket,
\end{equation}
where $e<0$ is the electron charge, and, again, $\hbar=c=1$. The presence of orbital magnetic moment of quasipaticles modifies the dispersion in band $n$ according to
\begin{equation}\label{eq:qpenergy}
  E_{n\pp}=\e_{n\pp}-\m_{n\pp}\BB.
\end{equation}

Denoting the renormalized band velocity with $\vv_{n\pp}=\p_\pp\e_{n\pp}-\p_\pp(\m_{n}\BB)$, the semiclassical equations of motion for band $n$ can be written as\cite{XiaoNiu}
\begin{eqnarray}\label{eq:EOM}
  \dot\rr&=&\vv_{n\pp}-\dot\pp\times{\bf \Omega}_{n\pp},\nonumber\\
  \dot\pp&=&e\EE+e\dot\rr\times\BB.
\end{eqnarray}
These equations yield
\begin{eqnarray}\label{eq:qpspeed}
  \dot\rr&=&\frac{1}{D_{\BB}}\left(\vv_{n\pp}-e\EE\times{\bf \Omega}_{n\pp}-e(\vv_{n\pp}\cdot{\bf \Omega}_{n\pp})\BB\right),\nonumber\\
  \dot\pp&=&\frac{1}{D_{\BB}}\left(e\EE+e\vv_{n\pp}\times\BB-e^2(\EE\cdot\BB){\bf \Omega}_{n\pp}\right),\nonumber\\
  D_\BB&=&1-e\BB{\bf \Omega}_{n\pp}.
\end{eqnarray}
In the equation for $\dot\rr$, the first term on the right hand side is the usual group velocity, including the effect of energy renormallization, Eq.~(\ref{eq:qpenergy}); the second one is the anomalous velocity~\cite{XiaoNiu}, associated with interband coherence effects induced by the electric part of the Lorentz force; the last contribution is the velocity that appears due to the interband coherence effects induced by the magnetic part of the Lorentz force, and is commonly associated with the static CME~\cite{Vilenkin,Nielsen1983,Cheianov1998,Kharzeev2008,Son2013,Niu2013}. The equation for $\dot\pp$, besides the usual Lorents force, contains an ``$\EE\cdot\BB$'' term, which is a manifestation of chiral anomaly at the quasiclassical level.\cite{SonSpivak2013,WongTserkovnyak2011} We note that the signs of the terms in the right hand sides of Eqs.~(\ref{eq:qpspeed}) vary between different works, which is related to the Berry curvature definition, and whether $e$ is taken to be positive or negative. We chose $e<0$ and ${\bf \Omega}_{n\pp}$ given by Eq.~(\ref{eq:Omega}). In accord with our plan to limit ourselves to linear response in $\EE$ and $\BB$ fields, in what follows we will disregard terms in Eqs.~(\ref{eq:qpspeed}) that are non-linear in electromagnetic fields. In particular, the chiral anomaly will play no role in our discussion.

Equations~(\ref{eq:qpspeed}) allow to write down the semiclassical kinetic equation.  We concentrate on the collisionless regime, $\omega\tau\to \infty$, where $\tau$ is the shortest relaxation time, since it is sufficient to bring out the points we would like to make. Later we will introduce a finite $\tau$ phenomenologically to discuss dissipative phenomena. The kinetic equation has the following form:
\begin{equation}\label{eq:kineqgeneral}
  \p_t f_{n\pp}+\dot\rr\p_\rr f_{n\pp}+\dot\pp\p_\pp f_{n\pp}=0.
\end{equation}

Finally, we have to establish the expression for the electric current. It contains two contributions: one, $\jj_{qp}$, that comes from the wave packet velocity of Eq.~(\ref{eq:qpspeed}), and the other coming from the curl of quasiparticle orbital magnetization, $\jj_m$. The former can be obtained from the continuity equation for the electric charge implied by kinetic equation~(\ref{eq:kineqgeneral});\cite{SonSpivak2013,Son2013} the latter is given by
\begin{equation}\label{eq:mcurrent}
  \jj_m=\p_\rr\times\sum_{n}\int (d\pp)\m_{n\pp} f_{n\pp},
\end{equation}
where $(d\pp)\equiv d^3p/(2\pi)^3$. This expression is expected on physical grounds, and can be formally obtained from a more low-level semiclassical kinetic equation for the full density matrix, by considering interband coherences established by gradients of the intraband distribution function.\cite{WongTserkovnyak2011,PesinMacDonaldMonopole,Son2013,MishchenkoStarykh2014}

 To find $\jj_{qp}$, we note that the density of electric charge is given by\cite{XiaoNiu} (note the $D_\BB$ factor)
\begin{equation}
  \rho=e\int(d\pp)D_\BB f_{n\pp},
\end{equation}
and must satisfy
\begin{equation}\label{eq:continuity}
  \p_t\rho+\p_\rr\jj_{qp}=0.
\end{equation}
(we neglected the chiral anomaly as a non-linear effect, hence no anomalous divergence in this equation. Alternatively, we could have noticed that $\EE\cdot\BB=0$ for an electromagnetic wave.) Multiplying the kinetic equation by $D_\BB$, observing that $D_\BB\p_t f_{n\pp}=\p_t(D_\BB f_{n\pp})-f_{n\pp}\p_t D_\BB$, and that according to the Faraday's law $\p_t \BB=-\p_\rr\times \EE$, after simple manipulations we obtain that $\jj_{qp}$ is given by
\begin{widetext}
\begin{equation}\label{eq:qpcurrent}
  \jj_{qp}=e\sum_{n}\int (d\pp)\left(\p_\pp\e_{n\pp}-\p_\pp(\m_{n}\BB)-e\EE\times{\bf \Omega}_{n\pp}-e(\p_\pp\e_{n\pp}\cdot{\bf \Omega}_{n\pp})\BB\right) f_{n\pp}.
\end{equation}
Combining Eqs.~(\ref{eq:mcurrent}) and (\ref{eq:qpcurrent}), we obtain the final expression for the electric current:
\begin{equation}\label{eq:totalcurrent}
  \jj=e\sum_{n}\int (d\pp)\left(\p_\pp\e_{n\pp}-\p_\pp(\m_{n}\BB)-e\EE\times{\bf \Omega}_{n\pp}-e(\p_\pp\e_{n\pp}\cdot{\bf \Omega}_{n\pp})\BB\right) f_{n\pp}+\p_\rr\times\sum_{n}\int (d\pp)\m_{n\pp} f_{n\pp},
\end{equation}
\end{widetext}
 For our purpose of finding a linear response to the fields, the distribution function $f_{n\pp}$ should be replaced with the equilibrium one whenever it is multiplied by $\EE$ or $\BB$ in Eq.~(\ref{eq:totalcurrent}).

Since we are interested in linear response to electromagnetic fields, we can simplify the kinetic equation~(\ref{eq:kineqgeneral}) further. The kinetic equation for band $n$, in which we keep only the terms linear in $\EE$ or $\BB$ (such are time and space derivatives of the distribution function), has the usual  form:
\begin{equation}\label{eq:kineq}
  \p_tf_{n\pp}+\p_\pp\e_{n\pp}\p_\rr f_{n\pp}+e\EE\p_\pp f_{n\pp}^0=0.
\end{equation}
Here $f_{n\pp}^0$ is the equilibrium distribution function, and only the electric part of the Lorentz force enters, since $\vv\times\BB\cdot\p_\pp f_{n\pp}^0=0$. This simplicity has a price: The usual galvanomagnetic phenomena~\cite{AbrikosovBook}, as well as the effects related to the chiral anomaly are then beyond the scope of the present treatment. Nonetheless, Eqs.~(\ref{eq:totalcurrent}) and (\ref{eq:kineq}) are sufficient to describe linear electric current response to $\EE$ and $\BB$ fields that vary slowly in space and time.

\subsection{Gyrotropic current}\label{sec:dynamic}

We now calculate the response at finite frequency and wave vector, satisfying $\w\gg vq$, where $v$ is the relevant speed of electrons in the crystal. We thus restrict ourselves to linear order in the wave vector $q$. We will see that it is the orbital magnetic moment of quasiparticles that is responsible for low-frequency chiral magnetic effect in clean metals, as well as the spatial dispersion of their optical conductivity.

For harmonic perturbations, the solution of the kinetic equation~(\ref{eq:kineq}) for the non-equilibrium part of the distribution function is trivially found by switching to Fourier space:
\begin{equation}\label{eq:df}
  \delta f_{n\pp}=\frac{1}{i(\w-\qq\p_\pp\e_{n\pp})}e\EE\p_\pp f_{n\pp}^0.
\end{equation}
We emphasize that since we are considering response at frequencies that are high compared to inverse relaxation times of the system, the equilibrium distribution function is given by $f_{th}(\e_{n\pp})$, \emph{not} by $f_{th}(E_{n\pp})$ with total quasiparticle energy~(\ref{eq:qpenergy}). The quasiparticle velocity is still given by $\pp$-space gradient of Eq.~(\ref{eq:qpenergy}), since the variation of magnetic field in time is slow on quantum-mechanical time scales.

The current is obtained from Eq.~(\ref{eq:totalcurrent}), keeping in mind that the terms with  $\EE$ or $\BB$ fields present must contain the equilibrium distribution function, $f_{n\pp}^0=f_{th}(\e_{n\pp})$. Those without the fields have $\delta f_{n\pp}$ from Eq.~(\ref{eq:df}) entering, since they are nullified by $f_{th}(\e_{n\pp})$.

Keeping only non-zero terms up to the linear order in $q E/\w$ or $B$, and taking into account that the anomalous Hall effect current coming from the anomalous velocity vanishes in a time-reversal invariant system, we obtain
\begin{widetext}
\begin{equation}\label{eq:dynamiccurrent}
\jj(\qq,\w)=\sum_{n}\int (d\pp)\left(\frac{e^2}{i\w}\p_\pp\e_{n\pp} (\EE\p_\pp f_{n\pp})-e^2(\p_\pp\e_{n\pp}\cdot{\bf \Omega}_{n\pp})f_{n\pp}\BB -e\p_\pp(\m_{n\pp}\BB)f_{n\pp}+\frac{e}{\w}(\qq\times\m_{n\pp}) (\EE\p_\pp f_{n\pp})\right).
\end{equation}
\end{widetext}
This expression for the gyrotropic current is one of the central results of our work. All terms in Eq.~(\ref{eq:dynamiccurrent}) have clear physical meaning: the first term is the reactive current that exists in any metal in the collisionless regime; the second term represents the static chiral magnetic effect,\cite{noteonSCME} discussed in Appendix~\ref{sec:static}; the third and forth contributions appear due to the presence of orbital magnetic moment of quasiparticles, the former being due to the corresponding energy renormalization, Eq.~(\ref{eq:qpenergy}), and the latter representing the current due to non-equilibrium orbital magnetization. The current due to the velocity renomalization does not vanish because one has to use $f_{th}(\e_n)$ as the unperturbed distribution function, as explained above. It does vanish if response to a purely static magnetic field is sought (thus in the absence of electric field), where truly equilibrium distribution function $f_{th}(E_n)$ is established, see Appendix~\ref{sec:static}.

The last three terms in Eq.~(\ref{eq:dynamiccurrent}) represent the leading contributions to the gyrotropic current, $\jj_g$, in a metal. Using integration by parts to rewrite all terms as contributions from the Fermi surface, we obtain
\begin{widetext}
\begin{equation}\label{eq:gyrocurrent}
\jj_g(\qq,\w)=\sum_{n}\int (d\pp)\left(e^2\e_{n\pp}(\p_\pp f_{n\pp}\cdot{\bf \Omega}_{n\pp})\BB +e(\m_{n}\BB)\p_\pp f_{n\pp}+\frac{e}{\w}(\qq\times\m_{n\pp}) (\EE\p_\pp f_{n\pp})\right).
\end{equation}

Having the expression for the gyrotropic current, we can write down the expression for tensor $\lambda_{abc}$ that determines the spatial dispersion of conductivity, Eq.~(\ref{eq:conductivity}):
\begin{equation}\label{eq:lambda}
\lambda_{abc}=-\frac{e^2}{\w}\sum_{n}\int (d\pp)\left(\e_{n\pp}(\p_\pp f_{n\pp}\cdot{\bf \Omega}_{n\pp})\e_{abc}+\frac{1}{e}m_{nd}\p_a f_{n\pp}\e_{dbc}+\frac{1}{e} m_{nd}\p_b f_{n\pp}\e_{adc}\right),
\end{equation}
\end{widetext}
where $\e_{abc}$ is the Levi-Civita tensor, $m_{nd}$ is the $d^{\text{th}}$ Cartesian component of $\m_{n\pp}$ and summation over repeated indices is implied. 

Eq.~(\ref{eq:lambda}) is useful when one is interested in a contribution to the conductivity made by a particular region of quasimomentum space, \textit{e.g.} close to a particular Weyl point. If only the total contribution is sought, the first term on the right hand side of Eq.~(\ref{eq:lambda}) vanishes upon $\pp$-integration and $n$-summation, and one is left with
\begin{equation}\label{eq:lambdaglobal}
\lambda_{abc}=-\frac{e}{\w}\sum_{n}\int (d\pp)\left(m_{nd}\p_a f_{n\pp}\e_{dbc}+ m_{nd}\p_b f_{n\pp}\e_{adc}\right).
\end{equation}
We use the same symbol for $\lambda_{abc}$ given by Eqs.~(\ref{eq:lambda}) or (\ref{eq:lambdaglobal}), even though they are equivalent only if the integration is done over the entire Brillouin zone, and summation is over all bands; this does not seem to lead to a confusion.

The validity of the entire treatment presented here is limited to frequencies that are small compared to the typical band gaps encountered in the problem. Only under this condition we can neglect interband absorption, and the dynamics is adiabatic, hence can be reduced to a single-band kinetic equation. For instance, for a single Weyl point, the frequency of the electromagnetic wave has to be small compared to the doping level; In a two-band model of a metal with spin-split Fermi surfaces, the frequency has to be small compared to the spin splitting at the Fermi level.

We briefly comment that the absorptive counterpart of natural optical activity - the circular dichroism - comes from the imaginary part of $\lambda_{abc}$, and can be included phenomenologically by substituting\cite{HosurQi2015} $\w\to \w+i/\tau$ into Eq.~(\ref{eq:lambdaglobal}),\cite{noteontau} where $\tau$ is the relevant relaxation time. For $\omega\tau\ll1$ we obtain a purely dissipative current, which is characterized by $\jj\propto \dot\BB$ in an isotropic system. This current, appearing due to the existence of Berry curvature in the band structure, is the ``intrinsic'' analog of the same type of current found in Ref.~\onlinecite{Levitov}, which considered electron scattering on impurities lacking an inversion center.

Finally, we would like to point out that our results are in full accord with the Onsager relations for conductivity,~\cite{LL8,Melrose} which in the absence of magnetic order read $\sigma_{ab}(\w,\qq)=\sigma_{ba}(\w,-\qq)$. This implies that tensor $\lambda_{abc}$ satisfies
\begin{equation}
  \lambda_{abc}(\w)=-\lambda_{bac}(\w).
\end{equation}
This is clearly the case for Eqs.~(\ref{eq:lambda}) and (\ref{eq:lambdaglobal}). Further, the Hermitian (absorptive) part of the conductivity tensor is an even function of $\w$, and its anti-Hermitian (dispersive) part is an odd function of $\w$. We can thus conclude that in the absence of absorption $\lambda_{abc}(\w)$ is a real tensor that satisfies
\begin{equation}
\lambda_{abc}(\w)=-\lambda_{abc}(-\w).
\end{equation}
Again, this holds for Eqs.~(\ref{eq:lambda}) and (\ref{eq:lambdaglobal}), which were derived neglecting dissipative effects. The imaginary (absorptive) part of $\lambda_{abc}$, describing circular dichroism, is even in frequency, which also happens to be true for the aforementioned phenomenological substitution $\w\to \w+i/\tau$ in Eq.~(\ref{eq:lambdaglobal}).

\subsection{Gyrotropic tensor. Relation to previous works}

In what follows, we find it convenient to switch to the description of gyrotropy in terms of the gyrotropic tensor. To this end we note that the third-rank tensor $\lambda_{abc}$, antisymmetric with respect to the first pair of indices, is dual to a second-rank pseudotensor - the gyrotropic tensor - $g_{ab}$. Indeed, both tensors have nine independent components, and change sign under inversion. The relation between the two tensors is given by~\cite{LL8}
\begin{equation}
  \lambda_{abc}=\e_{abd}g_{dc}, \quad g_{cd}=\frac12 \e_{abc}\lambda_{abd}.
\end{equation}
Using Eq.~(\ref{eq:lambda}), we obtain
\begin{widetext}
\begin{equation}\label{eq:gyrotensor}
  g_{ab}=-\frac{e^2}{\w}\sum_{n}\int (d\pp)\e_{n\pp}(\p_\pp f_{n\pp}\cdot{\bf \Omega}_{n\pp})\delta_{ab}-\frac{e}{\w}\sum_{n}\int (d\pp)\m_{n\pp}\cdot\p_\pp f_{n\pp}\delta_{ab}+\frac{e}{\w}\sum_{n}\int (d\pp)m_{na}\p_b f_{n\pp}.
\end{equation}
\end{widetext}
As is well-known,\cite{Niu2013} for any band structure the first term in this expression is just a complicated way to write a zero, while, for instance, it does make a contribution for a single Weyl point. This term also has to be taken into account if one considers a non-equilibrium situation, e.g. a Weyl metal, in  which different Weyl points have different chemical potentials.

The trace of the gyrotropic tensor is another useful quantity one may consider. It is easier to calculate than the full gyrotropic tensor, and if non-zero, immediately signals that the gyrotropic tensor itself is nonzero. In system with point groups of relatively high symmetry (isotropic systems in particular), the trace of the gyrotropic tensor solely determines the magnitude of dynamic CME. The expression for this trace is
\begin{widetext}
\begin{equation}\label{eq:gyrotrace}
  \text{Tr}g=-\frac{3e^2}{\w}\sum_{n}\int (d\pp)\e_{n\pp}(\p_\pp f_{n\pp}\cdot{\bf \Omega}_{n\pp})-\frac{2e}{\w}\sum_{n}\int (d\pp)(\m_{n\pp}\cdot\p_\pp f_{n\pp}).
\end{equation}
\end{widetext}

There is another aspect in which the trace of the gyrotropic tensor appears to be a quantity of interest. In general, the conductivity tensor~(\ref{eq:conductivity}) implies a complicated relation between the direction of current flow and that of magnetic field. Upon switching to the description in terms of the gyrotropic tensor, one can rewrite the latter as
\begin{equation}
  g_{ab}=\frac13 \text{Tr}g\delta_{ab}+\delta g_{ab},\quad \text{Tr}\delta g=0.
\end{equation}
The $\delta_{ab}$ part of this expression is easily seen to give rise to a current flowing along the magnetic field, since the corresponding tensor $\lambda_{abc}\propto \e_{abc}$. In a sense, this is a ``robust'' contribution to CME, independent of the propagation direction of an electromagnetic wave, and its polarization direction. The other, traceless, part may also lead to a current component along the magnetic field, but it is ``fine-tuned'': the magnitude of such effect would depend on the details
of wave propagation though a crystal. We therefore concentrate on calculating $\text{Tr}g$ in what follows, aiming at providing
 insight into the circumstances under which ``robust'' CME (in the sense explained above) can be observed.

Equations~(\ref{eq:dynamiccurrent}),~(\ref{eq:lambda}), (\ref{eq:lambdaglobal}), (\ref{eq:gyrotensor}) and (\ref{eq:gyrotrace}) are the central results of this work. They do not coincide with results published in works on related subjects~\cite{OrensteinMoore2013,Goswami2014,HosurQi2015,Moore2015}, which were derived for general bandstructures, the same as ours. We do agree with the results of Refs.~\onlinecite{Kharzeev2008,Son2013} for the dynamic chiral magnetic effect associated with a single Weyl point with particle-hole symmetry. In Ref.~\onlinecite{Mineev} the Berry curvature and orbital magnetic moment of quasiparticles are not explicit, thus the comparison of results is difficult. However, we do qualitatively agree with the latter work in simple limits (see below), but have different prefactors in final expressions. For a particular case of a two-band model, our expressions can be shown to reduce to those of Ref.~\onlinecite{ChangYang2015}. The result for the gyrotropic current of Ref.~\onlinecite{ChenBurkov2013} was obtained for a specific model with broken time-reversal and inversion symmetries, having only two Weyl points. The result pertains to the limit of strong magnetic field, and lies outside of the applicability region of the present theory. We can confirm, however, the general conclusion that in an ``ideal'' Weyl metal, represented by a collection of strictly particle-hole symmetric Weyl points located, in general, at different energies there exists dynamic CME, whose magnitude is determined by the energy-space dipole moment of Berry monopoles associated with the Weyl points. It is further claimed in Ref.~\onlinecite{ChenBurkov2013} that the chiral magnetic effect is a topological property of Weyl metals, with essentially universal magnitude. Our results that are applicable to general band structures do not confirm that observation: It appears that the existence of Weyl nodes is not even needed for a metal to show dynamic CME. We defer the discussion of specific cases that illustrate our claims till the next Section.

As far as we understand, the difference in results of the present work, and Refs.~\onlinecite{OrensteinMoore2013,HosurQi2015,Moore2015} is due to the fact that in the latter ones the magnetic part of the Lorentz force and the orbital magnetic moment of quasiparticles, Eq.~(\ref{eq:magneticmoment}), were not included in the semiclassical equations of motion, Eqs.~(\ref{eq:EOM}). As follows from our discussion, natural optical activity comes solely from these terms, and Refs.~\onlinecite{OrensteinMoore2013,HosurQi2015,Moore2015} should have obtained exact zero for, say, the gyrotropic tensor. The fact that finite results were obtained in these works can be traced back to their treatment of the current contribution coming from the anomalous velocity, which will denote $\jj_a$ for the time being. The corresponding expression is given by the third term in the first bracket on the right hand side of Eq.~(\ref{eq:totalcurrent}):
\begin{equation}
  \jj_a=-e^2\sum_{n}\int (d\pp)\EE\times{\bf \Omega}_{n\pp} f_{n\pp}.
\end{equation}
As we can see, this is a completely local current. Possible non-locality can only come from lattice constant scales, where the semiclassical description breaks down. Therefore, even for space-dependent $\EE(\rr)$ one gets zero for it to linear order in the electric field in a system with time reversal symmetry. The latter is due to the fact that time reversal symmetry implies that the Berry curvature is an odd function of $\pp$, ${\bf \Omega}_{n\pp}=-{\bf \Omega}_{n,-\pp}$, while the equilibrium distribution function is even. Further, the form of $\jj_a$ was essential to get the continuity equation~(\ref{eq:continuity}) to hold. Therefore, using alternative non-local expressions from Refs.~\onlinecite{OrensteinMoore2013,HosurQi2015,Moore2015} would violate charge conservation at the semiclasical level.

We would also like to point out that the  the gyrotropic tensor of Eq.~(\ref{eq:gyrotensor}) is written as a Fermi surface property, and is proportional to $1/\w$. This is a distinct feature of a metallic system: the leading frequency dependence of the gyrotropic tensor in an insulator is $\propto \w$.~\cite{Natori1975,Malashevich} The results of Refs.~\onlinecite{OrensteinMoore2013,Moore2015} also contain the $1/\w$ frequency dependence, yet cannot be written as Fermi surface contributions in general, and may give finite answers in band insulators; $\lambda_{abc}\propto1/\w$ in insulators contradicts Kramers-Kronig relations.~\cite{Malashevich}

In the remainder of this section we would like to discuss under what circumstances one should expect the trace of the gyrotropic not to vanish, and thus obtain robust dynamic CME. We note that the expression for the intrinsic orbital magnetic moment~(\ref{eq:magneticmoment}) can be rewritten as
\begin{eqnarray}\label{eq:magneticmomentdecomposition}
\m_{n\pp}&=&-e\e_{n\pp}{\bf \Omega}_{n\pp}+\nonumber\\
&&\frac {ie}2\sum_{m\neq n} (\e_m+\e_{n\pp})\bra \p_\pp u_{n\pp}|u_{m\pp}\ket\times\bra u_{m\pp}|\p_\pp u_{n\pp}\ket.\nonumber\\
&&
\end{eqnarray}
If one simply drops the second term on the right hand side of this expression, and plugs the first one into Eq.~(\ref{eq:gyrotrace}), the latter, after partial cancellations, will turn into an expression that determines the magnitude of the static CME, discussed in detail in Appendix~\ref{sec:static} (see Eq.~(\ref{eq:staticCME})). It is known that such an expression vanishes in any crystal.\cite{Niu2013} However,  it is in general \emph{{not correct}} to neglect that contribution. Moreover, in many important situations that term is the only source of non-trivial physical effects. For instance, in noncentrosymmetric Weyl metals, it is the presence of the second term in the right hand side of~(\ref{eq:magneticmomentdecomposition}) that gives a finite dynamic CME.

To determine when it is important to use the entire expression~(\ref{eq:magneticmomentdecomposition}) for the orbital magnetic moment, let us restrict ourselves to the case of only two bands coming close together in energy space in some region of the Brillouin zone, and assume that the sum over the intermediate states in Eq.~(\ref{eq:magneticmomentdecomposition}) is saturated by entries from these two bands. Then we can write that contribution to the magnetic moment for band $n$ as
\begin{widetext}
\begin{equation}\label{eq:deltam}
\frac {ie}2\sum_{m\neq n} (\e_m+\e_{n\pp})\bra \p_\pp u_{n\pp}|u_{m\pp}\ket\times\bra u_{m\pp}|\p_\pp u_{n\pp}\ket=
e\frac {(\e_{-n\pp}+\e_{n\pp})}2{\bf \Omega}_{n\pp},
\end{equation}
\end{widetext}
where we denoted the other band with $m=-n$. Clearly, deviations from exact particle-hole symmetry for the two bands in question will lead to the right hand side of Eq.~(\ref{eq:deltam}) being finite. A simple situation in which particle-hole asymmetry leads to the only finite contribution to the gyrotropic tensor of a noncentrosymmetric metal is presented in Section~\ref{sec:metal}.

\begin{figure}
  \centering
  \includegraphics[width=3 in]{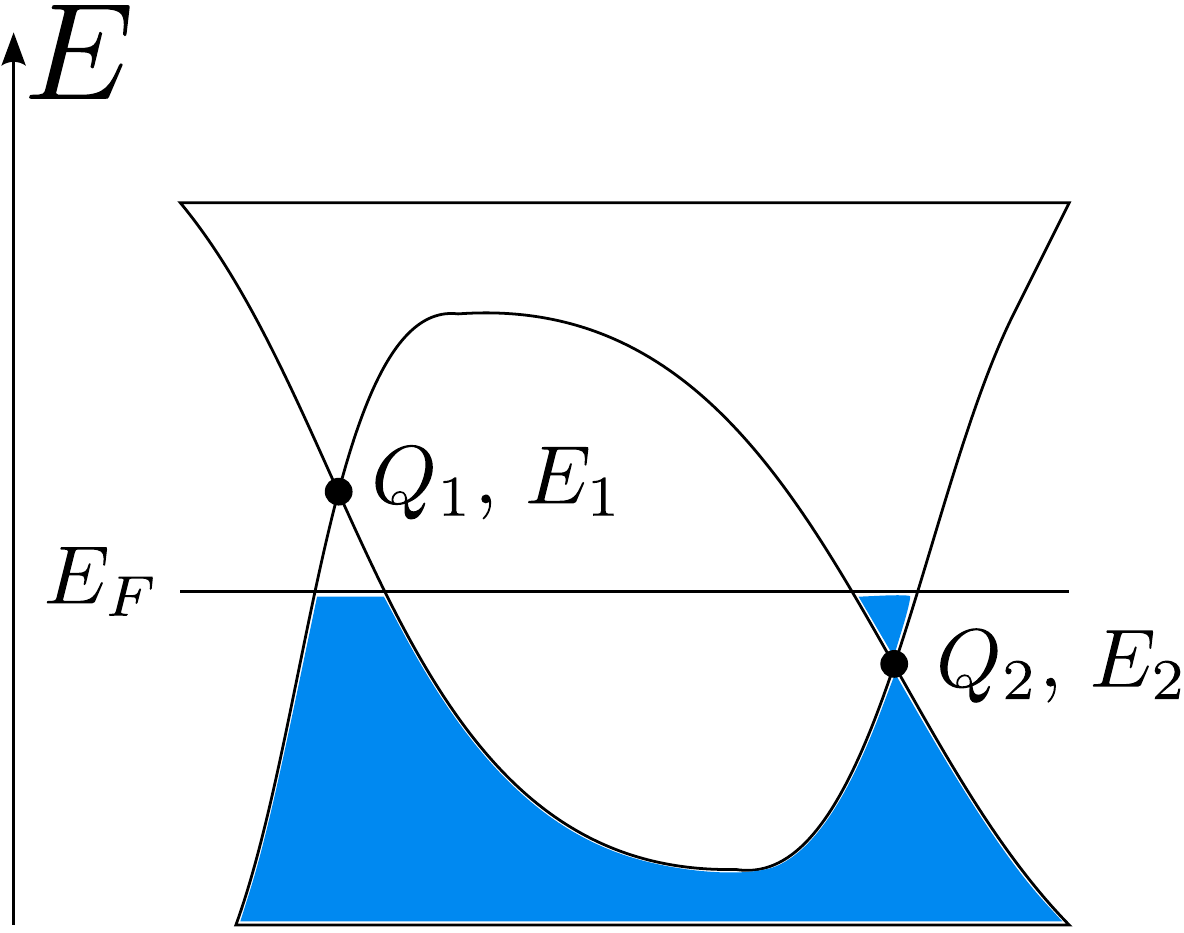}
  \caption{Schematic representation of a pair of Weyl points located at different energies close to the Fermi level of a material. $Q_{1,2}$ and $E_{1,2}$ are their chiralities and energies, respectively. The specific position of the Fermi energy $E_F$ with respect to the energies of the Weyl points (here chosen to be in between) is not important.}\label{fig:weylpoints}
\end{figure}

To proceed, we observe that the band energies that enter into Eq.~(\ref{eq:deltam}) are ``absolute'': The energy origin can be chosen arbitrarily, but it has to be the same for the entire band structure. Consider now a Weyl point with chirality $Q_w=\pm1$ located at energy $E_w$ near the Fermi level of a certain band structure, Fig.\ref{fig:weylpoints}. Even if one assumes perfect particle-hole symmetry around $E_w$ for this Weyl point, the right hand side of Eq.~(\ref{eq:deltam}) still does not vanish: $(\e_{-n}+\e_{n\pp})/2=E_w$. Since the latter expression is multiplied by the Berry curvature in Eq.~(\ref{eq:deltam}), and Berry curvature has a monopole-like  singularity near a Weyl point, the part of magnetic moment from Eq.~(\ref{eq:deltam}) will make a contribution to the total gyrotropic tensor that is proportional to $Q_wE_w$. It should be now clear that if one has a collection of Weyl points (of course, with zero net chirality), located at different energies, there total contribution to the current does not vanish as long as $\sum_wQ_wE_w$ does not vanish ($w$ now labels different Weyl points). The case where such a ``Berry dipole moment'' in energy space does not vanish is precisely when the dynamic CME in the system is expected~\cite{ChenBurkov2013}. This case is treated in more detail in Section~(\ref{sec:Weyl}).

Finally, it is not \textit{a priori} clear that the gyrotropic tensor and its trace require Berry monopoles to be non-zero. In fact, they do not, and in Section~\ref{sec:C4v} we provide an example of a situation where dynamic CME exists in a system without Weyl points.
\section{Natural optical activity and chiral magnetic effect in simple models}\label{sec:examples}

In this Section we will illustrate the expressions obtained above with specific examples of two-band models. We would like to use these as a platform to discuss under what circumstances CME appears in metallic systems.

The trace of the gyrotropic tensor~(\ref{eq:gyrotensor}) for a two-band system (TBS) is evaluated to be
\begin{equation}
  \text{Tr}g^{TBS}=-\frac{e^2}{\w}\sum_{n=\pm}\int (d\pp)\e_{-n}(\p_\pp f_{n\pp}\cdot{\bf \Omega}_{n\pp}).
\end{equation}
We used $\sum_{n=\pm}\int (d\pp)\e_{n}(\p_\pp f_{n\pp}\cdot{\bf \Omega}_{n\pp})=0$ to write this expression. The general momentum-space Hamiltonian for such a system is
\begin{equation}\label{eq:TBS}
  h^{TBS}={\boldsymbol \sigma}{\bf d}_\pp+E_\pp,
\end{equation}
where ${\boldsymbol \sigma}$ is a vector of Pauli matrices in the appropriate space (being the spin space in the examples to follow). The band energies are given by $\e_{n=\pm}=E_\pp\pm d_\pp$. Since $\e_{-n}=2E_p-\e_{n\pp}$, the trace of the gyrotropic tensor can be rewritten as
\begin{equation}\label{eq:TrTBS}
  \text{Tr}g^{TBS}=-\frac{2e^2}{\w}\sum_{n=\pm}\int (d\pp)E_\pp(\p_\pp f_{n\pp}\cdot{\bf \Omega}_{n\pp}).
\end{equation}
There is no general reason for this quantity to vanish. Expression~(\ref{eq:TrTBS}) can be used to make a few useful observations regarding the circumstances under which $\text{Tr}g$ does not vanish.

There are two distinct situation one may encounter evaluating Eq.~(\ref{eq:TrTBS}): $i$) $E_\pp$ takes a constant value on iso-energetic surfaces of $\e_{n\pp}$, $\e_{n\pp}=const$, and $ii$) it does not.

In case ($i$), the integral in Eq.~(\ref{eq:TrTBS}) can be split into that over magnitude of $\e_{n\pp}$, and a two dimensional surface integral over iso-energetic surfaces of $\e_{n\pp}$:
\begin{equation}
  \text{Tr}g^{TBS}=-\frac{2e^2}{\w}\sum_{n=\pm}\int d\e E_\pp(\e_{n\pp}=\e)\p_\e f_{n\pp}(\e)\int_{\e_{n\pp}=\e} d{\bf S}\cdot{\bf \Omega}_{n\pp}.
\end{equation}
Here $E_\pp(\e_{n\pp}=\e)$ denotes the value of $E_\pp$ on the surface $\e_{n\pp}=\e$, which is the same on the entire surface by definition in case ($i$). The surface integral then counts the total charge of Berry monopoles inside the $\e_{n\pp}=\e$ surface (Fermi surface for $\e=\mu$ at zero temperature). Hence, in case ($i$) in order to find a non-zero $\text{Tr}g$ one must have Berry monopoles in the band structure in the first place, and then hope that the energy integral and summation over bands do not render $\text{Tr}g$ zero. Sections \ref{sec:metal} and \ref{sec:Weyl} present examples where case ($i$) is realized, and it so happens that $\text{Tr}g\neq 0$.

Case ($ii$) is interesting when ${\bf \Omega}_{n\pp}$ is monopole-free. Here, even though $\int_{\e_{n\pp}=\e} d{\bf S}\cdot{\bf \Omega}_{n\pp}=0$, the fact that $E_\pp$ evaluates to different values on iso-energetic surfaces of $\e_{n\pp}$ can disrupt the exact cancellation between Berry fluxes over a closed surface, and yield a non-zero result for $\text{Tr}g$, and hence for dynamic CME, even without any Berry monopoles in the band structure. An example of such a situation is considered in Section~\ref{sec:C4v}.

\subsection{Isotropic noncentrosymmetric metal}\label{sec:metal}
The simplest possible model showing optical activity is that of an isotropic noncentrosymmetric metal, with the single particle Hamiltonian given by~\cite{Samokhin2007}
\begin{equation}\label{eq:brokenph}
  h_\pp=\frac{\pp^2}{2m}-\mu+v{\boldsymbol \sigma}\pp,
\end{equation}
where $m$ is the effective mass, and ${\boldsymbol \sigma}$ is a vector of Pauli matrices operating in the spin space. Loosely speaking, this model relates to a single Weyl point the same way as the Hamiltonian of a two-dimensional electron gas with Rashba spin-orbit coupling relates to that of a surface of a three-dimensional topological insulator: If only small $\pp$'s are considered, it describes a Weyl point with particle-hole symmetry breaking due to the mass term; if all $\pp$'s are taken into account, there are always two Fermi surfaces with exactly opposite Berry fluxes through them. Since we are considering this example essentially for the purpose of illustration of our results, we restrict ourselves to the case of zero temperature, and $\mu>0$, where each of the two bands has a Fermi surface.

The Hamiltonian~(\ref{eq:brokenph}) breaks all reflections (and thus the inversion), and symmetry-wise should allow natural optical activity. However, Refs.~\onlinecite{OrensteinMoore2013,Moore2015} would predict zero for the latter, while we get a non-zero result. Ref.~\onlinecite{Mineev} reaches the same conclusion as we do, but there the overall value of the gyrotropic current at low frequencies is four times as big as ours.

Due to the isotropy of model~(\ref{eq:brokenph}), its gyrotropic tensor is given by
\begin{equation}
  g_{ab}=\frac{1}{3}\text{Tr}g \delta_{ab}.
\end{equation}
We will evaluate separately the contributions of the two spin-split bands to $\text{Tr}g$, Eq.~(\ref{eq:gyrotrace}). Having single-band results will allow us to discuss the role of particle-hole asymmetry for a true Weyl point, where there is a single Fermi surface surrounding a Berry monopole.

We label the bands of Hamiltonian of Eq.~(\ref{eq:brokenph}) with $n=\pm$, and the corresponding energies, Berry curvatures and orbital magnetic moments are given by
\begin{equation}
  \e{\pm}=\frac{\pp^2}{2m}\pm vp, \quad {\bf \Omega}_{\pm}=\mp \frac 12\frac{\hat\pp}{p^2}, \quad \m_{\pm}=\frac{e v}{2}\frac{\hat \pp}{p}.
\end{equation}
The Fermi momenta for the two bands, $p_{\pm}$, are found from
\begin{equation}\label{eq:Fermimomenta}
  \frac{p^2_{\pm}}{2m}\pm vp_{\pm}=\mu.
\end{equation}

Restricting ourselves to zero temerature, and evaluating trivial integrals in Eq.~(\ref{eq:gyrotrace}), we obtain the contributions of the two band to the total gyrotropic tensor, $g_{\pm}$, to be (we restrict ourself to the dynamic limit only)
\begin{equation}
  g_{\pm}=\mp\frac{e^2}{12\pi^2\w}\left(\mu+\frac{p_{\pm}^2}{m}\right)\delta_{ab}.
\end{equation}
We see that the residual mass term contributes differently to $g_{\pm}$, and leads to a non-zero total $g=g_++g_-$
\begin{equation}
  g=-\frac{e^2}{12\pi^2\w}\left(\frac{p_{+}^2}{m}-\frac{p_{-}^2}{m}\right)\delta_{ab}.
\end{equation}
This result can be translated into an expression for the current response to (oscillating) uniform magnetic field:
\begin{equation}\label{eq:gyrocurrentmetal}
  \jj_g=\frac{e^2}{12\pi^2}\left(\frac{p_{+}^2}{m}-\frac{p_{-}^2}{m}\right)\BB.
\end{equation}
Analogous expression, but of magnitude four times as big, was obtained for this model in Ref.~\onlinecite{Mineev}. Further, formally looking like famous chiral magnetic effect, Eq.~(\ref{eq:gyrocurrentmetal}) shows that there is no universality in magnitude of such an effect in the dynamic limit: It depends on the details of the band structure, like band curvature, etc. We can estimate when such correction are important. Assuming that band curvature is due to a residual mass term with a typical for narrow gap semiconductor $m\sim 0.1$, and taking a typical value for the Dirac speed in Weyl metals to be\cite{Behrends2015} $v\sim 0.1-0.5 \times 10^{6}$m/s, we see that the energy scale that controls the importance of band curvature corrections is $mv^2\sim 10-100$meV. For doping levels of this order or larger, band curvature corrections need to be taken into account for quantitative description of chiral magnetic effect in Weyl metals.

\subsection{Weyl metal with particle-hole symmetric Weyl points}\label{sec:Weyl}

Now we would like to get expressions for the gyrotropic tensor in a simplified model of a noncentrosymmetric Weyl metal. The minimal example that would mimic the behavior of $g_{ab}$ in a real crystal is the approximation of the band structure with a collection of Weyl points. The simplest low-energy Hamiltonian in the immediate vicinity of each Weyl point, labeled by index $w$, is
\begin{equation}\label{eq:Hweylpoint}
  h_{w}=E_w-\mu_w+Q_w v_w{\boldsymbol \sigma}\pp_w,
\end{equation}
where $\pp_w$ is the deviation in quasimomentum from the location of the Weyl point in quasimomentum space, $E_w$ is the location of the Weyl point in energy space, $\mu_w$ is the chemical potential near it (different $\mu_w$'s would correspond to a non-equilibrium situation), $Q_w=\pm 1$ denotes the chirality, satisfying $\sum_w Q_w=0$, and $v_w$ is the band speed near each point. An obvious aspect of $H_w$ - its being  isotropic around the location of the Weyl point - does not limit the generality of obtained results. However, it is important that the  ``${\boldsymbol \sigma}\cdot \pp$'' form of Hamiltonian (\ref{eq:Hweylpoint}) makes the spectrum particle-hole symmetric around the energy of the Weyl point; the case of broken particle-hole symmetry is considered earlier in this Section.

For the model of Eq.~(\ref{eq:Hweylpoint}), the band index takes on two values $n=\pm$ around each Weyl point, with the corresponding energies given by $\e_{w\pm}=E_w\pm v_wp_w$. The Berry curvature is given by ${\bf \Omega}_{w\pm}=\mp Q_w{\hat\pp}_w/2\pp_w^2$, where ${\hat \pp}_w$ is the unit vector along $\pp_w$; the orbital magnetic moment is $\m_{w\pm}=-ev_wp_w{\bf \Omega}_{w\pm}$. It is crucial that it is $\e_{w\pm}$ that enter the first term in Eq.~(\ref{eq:gyrotensor}), yet it is $\e_{w\pm}-E_w=\pm v_w p$ (band splitting, rather than the individual band energy) that determine the orbital magnetic moment.

First, consider the case of equilibrium linear response, $\mu_w=\mu$ for all $w$'s. From Eqs.~(\ref{eq:gyrotensor}) and (\ref{eq:gyrotrace}) it is clear that only the terms that involve the orbital magnetic moment contribute to $g_{ab}$, and one obtains
\begin{equation}\label{eq:gyrotensorWeyl}
  g_{ab}=-\frac{e^2}{6\pi^2\w}\delta_{ab}\sum_{w}Q_wE_w, \quad \text{Tr}g=-\frac{e^2}{2\pi^2\w}\sum_{w}Q_wE_w.
\end{equation}
We observe that the trace of the gyrotropic tensor is determined by the dipole moment in energy space of the Berry curvature monopoles. The gyrotropic tensor itself implies that there exists a chiral-magnetic-effect-looking-like current given by
\begin{equation}
  \jj_g=\frac{e^2}{6\pi^2}\BB\sum_{w}Q_wE_w.
\end{equation}

We now turn to a hypothetical non-equilibrium case, in which the chemical potentials in different valleys do not coincide. In practice, such a situation can be reached using a separate set of $\EE$ and $\BB$ fields, with $\EE\cdot \BB\neq 0$ to drive the Weyl points out of equilibrium using the chiral anomaly.\cite{HosurQi2015} We, however, would like not to pay too much attention to the practical side of this case, and simply use it to show that  Eqs.~(\ref{eq:gyrotensor}) and (\ref{eq:gyrotrace}) give results known in the literature for this situation. For simplicity, we assume that all Weyl points are at the same energy $E_w=0$. Under non-equilibrium conditions, one should consider both static and dynamic limits. The former is analyzed based on the expression for the current of Eq.~(\ref{eq:chiralcurrent}), which corresponds to $g_{ab}$ from Eq.~(\ref{eq:gyrotensor}) with only the first term on the right hand side retained. Generalizing Eq.~(\ref{eq:currentcancellation}) to the case of Weyl points with different chemical potentials, one obtains
\begin{equation}\label{eq:gyrotensorWeylmu}
  g_{ab}=-\frac{e^2}{4\pi^2\w}\delta_{ab}\sum_{w}Q_w\mu_w,
\end{equation}
with the corresponding expression for the current,
\begin{equation}
  \jj_g=\frac{e^2}{4\pi^2}\BB\sum_{w}Q_w\mu_w.
\end{equation}
This is the original expression for the static chiral magnetic effect, first obtained by Vilenkin~\cite{Vilenkin}, and many times rederived after that.

In the dynamic limit, all terms in Eq.~(\ref{eq:gyrotensor}) do contribute to $g_{ab}$, and we obtain
\begin{equation}\label{eq:gyrotensorWeylmudynamic}
  g_{ab}=-\frac{e^2}{12\pi^2\w}\delta_{ab}\sum_{w}Q_w\mu_w,
\end{equation}
with the corresponding expression for the current,
\begin{equation}
  \jj_g=\frac{e^2}{12\pi^2}\BB\sum_{w}Q_w\mu_w.
\end{equation}
This is in full correspondence with ``chiral magnetic conductivity'' results of Ref.~\onlinecite{Kharzeev2008}, and results of Ref.~\onlinecite{Son2013}.

We summarize the outcomes of this subsection in Table~\ref{tab:cmc}, where we list the answers for the chiral magnetic conductivity, given by the ratio $j_g/B$, for the equilibrium case with Weyl points located at different energies, and for the non-equilibrium case with Weyl points located at different energies, for both static and dynamic situations. In Table~\ref{tab:cmc}, the only non-zero entry of topological origin is the static chiral conductivity in the non-equilibrium case. The universal flavor of the rest of results is misleading, it is just a consequence of the simplified model of Eq.~(\ref{eq:Hweylpoint}). The entries of third column of the Table can be found in the literature. To the best of our knowledge, the bottom entry of the second column is obtained here for the first time.
\begin{center}
\begin{table}
\begin{tabular}{|c|c|c|}
  \hline
     & $\mu_w=\mu$ & $E_w=E$ \\
  \hline
  Static & 0 & $\frac{e^2}{4\pi^2}\sum_{w}Q_w\mu_w$ \\
  \hline
  Dynamic & $\frac{e^2}{6\pi^2}\sum_{w}Q_w E_w$ & $\frac{e^2}{12\pi^2}\sum_{w}Q_w\mu_w$ \\
  \hline
\end{tabular}
\caption{Chiral magnetic conductivity for a Weyl metal for different cases of linear response theory. The rows correspond to static and dynamic limits of response. The columns pertain to the equilibrium response of a Weyl metal with Weyl points located at different energies, and to non-equilibrium response of a Weyl metal with Weyl points at the same energy.}
\label{tab:cmc}
\end{table}
\end{center}

\subsection{Chiral magnetic effect without Berry monopoles}\label{sec:C4v}

In Sections~\ref{sec:metal} and \ref{sec:Weyl} we considered examples of Hamiltonians for noncentrosymmetric metals that lead to CME. In both cases it was the presence of linear band tounchings - Weyl points - that was the culprit, even though the example of Section~\ref{sec:metal} would not be ordinarily called a Weyl metal. Is the existence of Weyl points, or rather of Berry curvature monopoles, necessary for the existence of the chiral magnetic effect? In this Section we show that the answer to that question is ``no''.

To show that Berry monopoles are not required for the chiral magnetic effect, we provide an explicit example of a Hamiltonian for which the Berry curvature is monopole-free, yet it exhibits CME. We start with the simplest spin-orbit coupling Hamiltonian for a metal with noncentrosymmetric point group $C_{4v}$, given by Eq.~(\ref{eq:TBS}) with the following ${\bf ' d}_\pp$:~\cite{SamokhinMineev2008}
\begin{equation}
  {\bf d}^{C_{4v}}_\pp=(v p_y,-vp_x,\gamma p_xp_yp_z(p_x^2-p_y^2)).
\end{equation}
This Hamiltonian is symmetric under the four-fold rotation around the $z$-axis, breaks $z\to-z$ reflection, but preserves $x\to-x$ and $y\to -y$ reflections, as appropriate for $C_{4v}$ point group. The presence of $x\to-x$ and $y\to -y$ reflections makes this group not gyrotropic (it does not show natural optical activity). To break these reflections, we choose $E_\pp$ in Eq.~(\ref{eq:TBS}) as
\begin{equation}
  E^{C_4}_{\pp}=\frac{p^2}{2m}+\gamma_vp_xp_y(p_x^2-p_y^2),
\end{equation}
in which the second term does the job, keeping the four-fold rotation around $z$ intact. The total Hamiltonian thus has gyrotropic $C_4$ symmetry,
\begin{equation}\label{eq:C4}
  h^{C_4}=\frac{p^2}{2m}+\gamma_vp_xp_y(p_x^2-p_y^2)+{\boldsymbol \sigma}{\bf d}^{C_{4v}}_\pp,
\end{equation}
and we expect that $\gamma_v$ determines the optical activity properties of this model.

The Berry curvature in the band structure that corresponds to Hamiltonian~(\ref{eq:C4}) is determined by the ``parent'' $C_{4v}$ theory:
\begin{equation}
  {\bf \Omega}_{\pm}=\mp\gamma v^2\frac{p_xp_y(p_x^2-p_y^2)}{2(d^{C_{4v}}_\pp)^3}(p_x,p_y,-3p_z).
\end{equation}
This Berry curvature is obviously monopole-free, it is not even singular near the origin. To calculate $\text{Tr}g$ in this case, we set temperature to zero, and assume that the spin splitting of the band structure and breaking of $x\to-x$ and $y\to-y$ reflections are small, $\gamma p_F^5,\gamma_v p_F^4\ll\mu$, where the Fermi momentum $p_F$ is found from $p_F^2/2m=\mu$. To evaluate Eq.~(\ref{eq:TrTBS}), we integrate it by parts to move differentiation with respect to $\pp$ onto $E_\pp$ ($\p_\pp{\bf \Omega}_{n\pp}=0$ in this case), and expand the distribution function to linear order in $d_\pp$ and $\gamma_v$:
\begin{widetext}
\begin{equation}\label{eq:Fexpansion}
  f_{\pm}(\e_{\pm})\approx f_{th}(p^2/2m-\mu)-\p_\mu f_{th}(p^2/2m-\mu)(\gamma_vp_xp_y(p_x^2-p_y^2)\pm d_\pp).
\end{equation}
Only the second term on the right hand side of Eq.~(\ref{eq:Fexpansion}) contributes to $\text{Tr}g$. In addition, the derivative of the Fermi function in that term restricts the integration over momenta to $p=p_F$ at zero temperature. After simple transformations, we obtain
\begin{equation}
  \text{Tr}g=\frac{e^2}{\w}\frac{vp_F\cdot\gamma_vp_F^4}{\mu}I\left(\frac{\gamma p_F^4}{v}\right),
\end{equation}
where
\begin{equation}
  I(x)=\frac{1}{2\pi^3}\int^{2\pi}_0 d\phi\int^\pi_0 d\theta \sin\theta\frac{x}{x^2\cos^2\theta+16\sin^{-6}\theta\sin^{-2}(4\phi)}.
\end{equation}
\end{widetext}
It is easy to find asymptotic values of this integral:
\begin{equation}
  I(x)\approx\begin{cases}
    \frac{x}{35\pi^2}, & \text{if $x\ll 1$},\\
    \frac{1}{2\pi^2}, & \text{if $x\gg1$}.
  \end{cases}
\end{equation}
This non-zero result for the trace of the gyrotropic tensor illustrates the main point of this Section: Berry monopoles are not needed for the dynamic chiral magnetic effect. In particular, the dynamic chiral magnetic effect is not related to the chiral anomaly in general.
\section{Conclusion}\label{sec:conclusion}

To summarize, we have derived a completely general expressions (Eqs.~(\ref{eq:chiralcurrent}) and \ref{eq:dynamiccurrent})) for the leading contribution to the gyrotropic current in a metallic system in the limit of small frequencies and wave vectors: $\w,vq\ll \mu, E_g$, where $E_g$ is the relevant energy gap, and $v$ is the relevant speed. The typical doping levels found in newly discovered Weyl metals range from 1meV to 10meV, which places $\w$ in the THz range. To confirm our results, derived from the semiclassical kinetic equation, we performed a calculation based on the Kubo formula, which yielded identical ones.

The main physical conclusion is the identification of the intrinsic orbital magnetic moment of quasiparticles as the source of natural optical activity in (semi)metals. We also showed that the chiral magnetic effect in general exists in the dynamic limit ($\w>vq$) in metallic systems with natural optical activity. The Weyl metal with a gyrotropic point group is such a system. However, the list of band structures showing natural optical activity (and chiral magnetic effect in particular) is not limited to the Weyl metal: The presence of Weyl points and the associated Berry monopoles is in general not required for the existence of chiral magnetic effect and natural optical activity at low frequencies. This was also numerically confirmed in a recent preprint.~\cite{ChangYangpreprint}

Specializing to the case of Weyl metals, if the latter can be represented by a collection of particle-hole symmetric Weyl points near its Fermi level, the trace of the gyrotropic tensor, determining the magnitude of the dynamic CME, is determined by the energy space dipole moment of Berry monopoles, corresponding to the Weyl points. In general, this trace, as well as the magnitude of the chiral magnetic effect, is determined by peculiarities of the band structure, and is non-universal; We estimate the doping level above which band curvature effects are important to take into account in description of dynamic CME to be $10-100$ meV.

Recently discovered\cite{Huang2015,Lv2015,Shekhar2015,Shekhar2015n2} Weyl metals from mono-pnictide family\cite{BernevigDai2015} have $C_{4v}$ point group, which is not gyrotropic.\cite{Nye} One could hope to study natural optical activity in these materials upon application of an appropriate strain to reduce the point group to, say, $C_4$. Even in that case such a study would be complicated by their opaqueness to due high concentrations of mobile carriers. This obstacle may be possible to overcome with applying pressure, which has been reported to drive TaAs toward insulating behavior\cite{Zhou2015}

\begin{acknowledgements}
We are grateful to Boris Spivak for asking us questions that led to starting this project. We also thank Andrei Malashevich and Joel Moore for useful discussions. This work was supported by NSF Grant No. DMR-1409089
\end{acknowledgements}

\appendix
\begin{widetext}

\section{Derivation of (\ref{eq:gyrocurrent}) from Kubo formula}
In this section we rederive the expressions for the gyrotropic current, obtained above from the semiclassical kinetic equation, using Kubo formula. Given the spread of results existing in the literature, we would like to give all the details of the calculation, such that it would be straightforward (but unavoidably time consuming) to check it.

We adopt the gauge in which the electric potential is equal to zero, and seek current response to vector potential, ${\bf A}$. The Kubo formula for the Fourier component of the current relates the latter to the vector potential, $j_a(\w,\qq)=Q_{ab}A_b(\qq,\w)$, with kernel $Q_{ab}$ given by
\begin{equation}\label{eq:Q}
Q_{ab}(\w, \qq)=e^2\sum_{n,n^{\prime}}\int (d\pp)\frac{f_{n,\pp+\qq/2}-f_{n^{\prime},\pp-\qq/2}}
{\w-\xi_{n,\pp+\qq/2}+\xi_{n^{\prime},\pp-\qq/2}}
\bra u_{n^{\prime},\pp-\qq/2}|\p_a h_{\pp}|u_{n,\pp+\qq/2}\ket\bra u_{n,\pp+\qq/2}|\p_b h_{\pp}|u_{n^{\prime},\pp-\qq/2}\ket.
\end{equation}
We will only consider the linear-in-$\qq$ part of $Q_{ab}$, since the $O(\qq^0)$ terms are completely standard. Normally, in order to compute the physical conductivity one has to subtract the diamagnetic current contribution, which amounts to redefining $Q$ according to $Q(\w,\qq)\to Q(\w,\qq)-\lim_{\qq\to 0}Q(0,\qq)$. Since we are interested in $O(q)$ part of $Q$, we do not have to worry about the diamagnetic term.

We will split the total response into the intra- and interband parts, starting with the intraband one.

\subsection{Intraband part}

For clarity and brevity, we will use $|u_{n}\ket\equiv|u_{n\pp}\ket$, $\e_n\equiv \e_{n\pp}$, $h\equiv h_\pp$, and $f_n\equiv f_{n\pp}$. For band velocity we will use $\p_\pp\e_{n\pp}\equiv\vv_{n\pp}$, and the Cartesian components of $\vv_{n\pp}$ will be denoted with letters from the beginning of Latin alphabet: $v_{nc}$ denotes the $c^{\text{th}}$ component of velocity in band $n$.

Linear-in-q intraband contribution exists only in the static limit of Eq.~(\ref{eq:Q}) ($\w\to0$ before $\qq\to0$.) The contribution to $Q_{ab}$ that appears in the static limit we denote with $Q_{ab}^{intra}(0, \qq)$.

\begin{eqnarray}
Q_{ab}^{intra}(0, \qq)&=&-e^2\sum_{n}\int(d\pp)
\frac{f_{n\pp +\qq/2}-f_{n\pp -\qq/2}}
{\xi_{n\pp +\qq/2}-\xi_{n\pp -\qq/2}}
\bra u_{n\pp -\qq/2}|\p_a h_{\pp }|u_{n\pp +\qq/2}\ket
\bra u_{n\pp +\qq/2}|\p_b h_{\pp}|u_{n\pp -\qq/2}\ket=\nonumber\\
&=&-e^2\sum_{n}\int(d\pp)
\frac{\vv_{n\pp}\qq}{\vv_{n\pp}\qq}\p_{\e_{n\pp}} f_{n\pp}
\bra u_{n\pp -\qq/2}|\p_a h_{\pp }|u_{n\pp +\qq/2}\ket
\bra u_{n\pp +\qq/2}|\p_b h_{\pp }|u_{n\pp -\qq/2}\ket.
\end{eqnarray}
We see that the entire response comes from the matrix elements of the velocity, which we need to expand to first order in $\qq$. Consider the first bracket:
\begin{eqnarray}
\bra u_{n\pp-\qq/2}|\p_a h|u_{n\pp+\qq/2}\ket&=&\bra u_{n}|\p_a h|u_{n}\ket-\frac12 q_c\bra\p_c u_{n}|\p_ah|u_{n}\ket+\frac12 q_c\bra u_{n}|\p_ah|\p_c u_{n}\ket=\nonumber\\
&=&\p_a \e_{n}-\frac12 q_c\bra\p_c u_{n}|\p_ah|u_{n}\ket+\frac12 q_c\bra u_{n}|\p_ah|\p_c u_{n}\ket.
\end{eqnarray}
To get the second bracket, we have to change $a\to b$, and $\qq\to -\qq$:
\begin{eqnarray}
&&\bra u_{n\pp+\qq/2}|\p_b h|u_{n\pp-\qq/2}\ket=\p_b \e_{n}+\frac12 q_c\bra\p_c u_{n}|\p_b h|u_{n}\ket-\frac12 q_c\bra u_{n}|\p_b h|\p_c u_{n}\ket.
\end{eqnarray}
Then we get
\begin{eqnarray}
Q_{ab}^{intra}(0, \qq)&=&-e^2\sum_{n}\int(d\pp)\p_{\e_{n}} f_{n}
\bra u_{n\pp-\qq/2}|\p_a h|u_{n\pp+\qq/2}\ket
\bra u_{n\pp+\qq/2}|\p_b h|u_{n\pp-\qq/2}\ket\approx\nonumber\\
&=&-\frac{e^2}2\sum_{n}\int(d\pp)\p_{\e_{n}} f_{n} q_c
\left(v_{na} \bra\p_c u_{n}|\p_b h|u_{n}\ket-v_{na} \bra u_{n}|\p_b h|\p_c u_{n}\ket-v_{nb} \bra\p_c u_{n}|\p_ah|u_{n}\ket+v_{nb} \bra u_{n}|\p_ah|\p_c u_{n}\ket\right)\nonumber\\
&&
\end{eqnarray}
Now in each term on the right hand side we insert a resolution of identity, $\sum_{n'}|u_{n'}\ket\bra u_{n'}|={\bf 1}$, between the derivative of a bra or a ket, and a derivative of the Hamiltonian, and observe that the terms with $n'=n$ cancel.

Further progress is possible if one uses the following identities:
\begin{eqnarray}\label{eq:identities}
&&\bra u_{n}|\p_ah|u_{n}\ket=\p_a\e_{n},\quad n= n'\nonumber\\
&&\bra u_{n}|\p_ah|u_{n'}\ket=(\e_{n'}-\e_{n})\bra u_{n}|\p_au_{n'}\ket,\quad n\neq n'\nonumber\\
&&\bra u_{n}|\p_au_{n'}\ket=-\bra \p_au_{n}|u_{n'}\ket
\end{eqnarray}
The first two of these can be compactly written as\cite{MalashevichVanderbilt2010}
\begin{equation}\label{eq:theidentity}
  \p_a(h_\pp-\e_{n})|u_{n}\ket=(\e_{n}-h_\pp)|\p_au_{n}\ket,
\end{equation}
which is the form we found most useful.

Using Eq.~(\ref{eq:theidentity}), we obtain
\begin{eqnarray}\label{eq:Q1}
Q_{ab}^{intra}(0, \qq)&=&-\frac{e^2}2\sum_{n'\neq n}\int(d\pp)\p_{\e_{n}} f_{n} q_c
v_{na}(\e_{n}-\e_{n'})\left( \bra\p_c u_{n}|u_{n'}\ket\bra u_{n'}|\p_bu_{n}\ket-\bra\p_bu_{n}|u_{n'}\ket \bra u_{n'}|\p_c u_{n}\ket\right)+\nonumber\\
&&+\frac{e^2}2\sum_{n'\neq n}\int(d\pp)\p_{\e_{n}} f_{n} q_c v_{nb}(\e_{n}-\e_{n'}) \left(\bra\p_c u_{n}|u_{n'}\ket\bra u_{n'}|\p_au_{n}\ket- \bra \p_a u_{n}| u_{n'}\ket\bra u_{n'}|\p_c u_{n}\ket\right).
\end{eqnarray}
This expression can be related to the quasiparticle orbital magnetic moment, given by
\begin{equation}
  \m_{n}=\frac{i e }{2}\bra\p_\pp u_{n}|\times(h_\pp-\e_{n})|\p_\pp u_{n}\ket,
\end{equation}
or in components:
\begin{equation}
  \m_{nd}=\frac{i e }{2}\e_{drs}\bra\p_r u_{n}|\times(h_\pp-\e_{n})|\p_s u_{n}\ket=-\frac{i e }{2}\sum_{n'\neq n}\e_{drs}(\e_{n}-\e_{n'})\bra\p_r u_{n}|u_{n'}\ket\bra u_{n'}|\p_s u_{n}\ket.
\end{equation}
From the last equation it follows that
\begin{equation}
  \e_{dab}\m_{nd}=-\frac{i e }{2}\sum_{u_{n'}\neq n}(\e_{n}-\e_{n'})(\bra\p_a u_{n}|u_{n'}\ket\bra u_{n'}|\p_bu_{n}\ket-\bra\p_b u_{n}|u_{n'}\ket\bra u_{n'}|\p_au_{n}\ket).
\end{equation}
This allows one to rewrite Eq.~(\ref{eq:Q1}) as
\begin{eqnarray}\label{eq:Q2}
Q_{ab}^{intra}(0, \qq)&=&-ie\sum_{n}\int(d\pp)\p_{\e_{n}} f_{n} q_c
v_{na}\e_{dcb}\m_{nd}+ie\sum_{n}\int(d\pp)\p_{\e_{n}} f_{n} q_c v_{nb}\e_{dca}\m_{nd}.
\end{eqnarray}
The corresponding contribution to the gyrotropic current is
\begin{equation}\label{eq:intracurrent}
  {\bf j}^{intra}_g=-e\sum_{n}\int(d\pp) \m_{n}\cdot(i\qq\times {\bf A})\p_{\pp}f_{n}-e\sum_{n}\int(d\pp) (i\qq\times\m_{n})({\p_{\pp} f_{n} \cdot{\bf A}}).
\end{equation}
These terms have very different semiclassical interpretation. The first one is the current that comes from energy shift in the semiclassical distribution function, Eq.~(\ref{eq:f}). There exists a related contribution that corresponds to band velocity renormalization due to energy renormalization, Eq.~(\ref{eq:qpenergy}), which is a part of the intraband response, see below. The second term is equal (with the opposite sign) to the current of non-equilibrium quasiparticle magnetization. Its role in the derivation is to cancel the former contribution, coming from the intraband part of response, in the static limit.

\subsection{Interband part}
The interband part of response is insensitive to the order of $\w\to0$ and $q\to 0$ limits. We will thus write
\begin{eqnarray}
Q_{ab}^{inter}(0, \qq)&=&-e^2\sum_{n}\int(d\pp)
\frac{f_{n\pp +\qq/2}-f_{n',\pp -\qq/2}}
{\e_{n\pp+\qq/2}-\e_{n'\pp-\qq/2}}
\bra u_{n',\pp -\qq/2}|\p_a h|u_{n\pp +\qq/2}\ket
\bra u_{n\pp +\qq/2}|\p_b h|u_{n',\pp -\qq/2}\ket.
\end{eqnarray}
There are three sources of linear-in-q terms: the distribution function difference, the difference of transition energies, and the matrix element. We consider them one by one below.
\subsubsection{Distribution function contribution}
The contribution to $Q_{ab}$ that comes from $q$-dependence of distribution functions difference for interband transitions we denote with $Q_{ab}^{inter,df}(0, \qq)$.
\begin{eqnarray}
Q_{ab}^{inter,df}(0, \qq)&=&-\frac12e^2\sum_{n}\int(d\pp)
\frac{\qq\p_\pp f_{n}+\qq\p_\pp f_{n'}}
{\e_{n}-\e_{n'}}
\bra u_{n'}|\p_a h|u_{n}\ket
\bra u_{n}|\p_b h|u_{n'}\ket=\nonumber\\
&=&-\frac12e^2\sum_{n}\int(d\pp)
(\qq\p_\pp f_{n}+\qq\p_\pp f_{n'})(\e_{n}-\e_{n'})
\bra \p_a u_{n'}|u_{n}\ket
\bra u_{n}|\p_b u_{n'}\ket=\nonumber\\
&=&-\frac12e^2\sum_{n}\int(d\pp)
(\qq\p_\pp f_{n})(\e_{n}-\e_{n'})
\left(\bra \p_b u_{n}|u_{n'}\ket\bra u_{n'}|\p_a u_{n}\ket-\bra \p_a u_{n}|u_{n'}\ket \bra u_{n'}|\p_b u_{n}\ket\right)=\nonumber\\
&=&-ie\sum_{n}\int(d\pp)
(\qq\p_\pp f_{n})\e_{cba}\m_{nl}.
\end{eqnarray}
The corresponding gyrotropic current is
\begin{equation}
  {\bf j}^{inter,df}_{g}=-ie\sum_{\pp,n}
(\qq\p_\pp f_{n})\m_{n}\times {\bf A}.
\end{equation}
We note in passing that
\begin{equation}\label{eq:currentdf}
  {\bf j}^{intra}_g+{\bf j}_{g}^{inter,df}=-\frac{e}{  c}\sum_{\pp,n}(\p_\pp f_{n}\m_{n})(i\qq\times A)=-\frac{e}{  c}\sum_{\pp,n}\p_{\e_{n}} f_{n}(\vv_{n}\m_{n})\BB.
\end{equation}

\subsubsection{Transition energies}
The contribution to $Q_{ab}$ that comes from $q$-dependence of transition energies for interband transitions we denote with $Q_{ab}^{inter,te}(0, \qq)$.
\begin{eqnarray}
Q_{ab}^{inter,te}(0, \qq)&=&e^2\sum_{n\neq n'}\int(d\pp)
\frac{f_{n}-f_{n'}}
{(\e_{n}-\e_{n'})^2}(v_{nc}+v_{n'c})\frac{q_c}{2}
(\e_{n}-\e_{n'})\bra u_{n'}|\p_a u_{n}\ket
(\e_{n'}-\e_{n})\bra u_{n}|\p_b u_{n'}\ket=\nonumber\\
&=&e^2\sum_{n\neq n'}\int(d\pp)
(f_{n}-f_{n'})(v_{nc}+v_{n'c})\frac{q_c}{2}
\bra u_{n'}|\p_a u_{n}\ket
\bra \p_b u_{n}| u_{n'}\ket
\end{eqnarray}
Because of $f_{n}-f_{n'}$ we can extend the summation to include $n=n'$. The terms that involve products like $f_{n}v_{n}$ and $f_{n'}v_{n'}$ allow summation over the intermediate states, such that we obtain
\begin{eqnarray}\label{eq:te1}
Q_{ab}^{inter,te}(0, \qq)&=&e^2\sum_{n}\int(d\pp)
f_{n}v_{nc}\frac{q_c}{2}\left(\bra \p_b u_{n}| \p_a u_{n}\ket-i\leftrightarrow j\right)+
e^2\sum_{n,n'}\int(d\pp)
f_{n}v_{n'c}\frac{q_c}{2}\left(\bra u_{n'}| \p_a u_{n}\ket\bra \p_b u_{n}|u_{n'}\ket-i\leftrightarrow j\right).\nonumber\\
&&
\end{eqnarray}
Using identity~(\ref{eq:theidentity}), we can do more summations over intermediate states in Eq.~(\ref{eq:te1}):
\begin{eqnarray}\label{eq:te2}
Q_{ab}^{inter,te}(0, \qq)&=&e^2\sum_{\pp,n}
f_{n}\frac{q_c}{2}\left(\bra \p_b u_{n}|\p_c(h+\e_{n})| \p_a u_{n}\ket-i\leftrightarrow j\right)-\nonumber\\
&&-e^2\sum_{\pp,n,n'}
f_{n}\frac{q_c}{2}\left(\bra u_{n'}|\p_a u_{n}\ket\bra\p_b u_{n} |(\e_{n'}-h)|\p_c u_{n'}\ket-i\leftrightarrow j\right).
\end{eqnarray}

\subsubsection{Matrix elements}
The contribution to $Q_{ab}$ that comes from $q$-dependence of matrix elements for interband transitions we denote with $Q_{ab}^{inter,me}(0, \qq)$.
\begin{eqnarray}
Q_{ab}^{inter,me}(0, \qq)&=&-e^2\sum_{\pp,n\neq n'}
\frac{f_{n}-f_{n'}}
{(\e_{n}-\e_{n'})}
\bra u_{n'\pp-\qq/2}|\p_ah|u_{n\pp+\qq/2}\ket
\bra u_{n\pp+\qq/2}|\p_b h|u_{n'\pp-\qq/2}\ket.
\end{eqnarray}

Using
$$\bra u_{n',\pp-\qq/2}|\p_ah|u_{n\pp+\qq/2}\ket=(\e_{n}-\e_{n'})\bra u_{n'}|\p_a u_{n}\ket-\frac{q_c}2\bra \p_c u_{n'}|\p_ah|u_{n}\ket+\frac{q_c}2\bra u_{n'}|\p_ah|\p_c u_{n}\ket,$$
$$\bra u_{n\pp+\qq/2}|\p_bh|u_{n',\pp-\qq/2}\ket=(\e_{n}-\e_{n'})\bra \p_b u_{n}|u_{n'}\ket+\frac{q_c}2\bra \p_c u_{n}|\p_bh|u_{n'}\ket-\frac{q_c}2\bra u_{n}|\p_bh|\p_c u_{n'}\ket$$
the linear-in-q part of the matrix element contribution becomes
\begin{eqnarray}\label{eq:me1}
Q_{ab}^{inter,me}(0, \qq)=-e^2\sum_{\pp,n,n'}
(f_{n}-f_{n'})\frac{q_c}2&&\left(\bra u_{n'}|\p_a u_{n}\ket\bra \p_c u_{n}|\p_b h|u_{n'}\ket-\bra u_{n'}|\p_a u_{n}\ket\bra u_{n}|\p_b h|\p_c u_{n'}\ket-\right.\nonumber\\
&&\left.-\bra \p_b u_{n}| u_{n'}\ket\bra \p_c u_{n'}|\p_a h|u_{n}\ket+\bra \p_b u_{n}| u_{n'}\ket\bra u_{n'}|\p_a h|\p_c u_{n}\ket\right).
\end{eqnarray}
Performing summation over intermediate states wherever possible, and exchanging $n\leftrightarrow n'$ where convenient, we obtain
\begin{eqnarray}\label{eq:me2}
Q_{ab}^{inter,me}(0, \qq)&=&e^2\sum_{n}\int(d\pp) f_{n}\frac{q_c}2(\bra\p_a u_{n}|\p_b h|\p_c u_{n}\ket+[cab]-[cba]-[bac])+\nonumber\\
&+&e^2\sum_{n,n'}\int(d\pp) f_{n}\frac{q_c}2\left( \bra u_{n'}|\p_a u_{n}\ket\bra u_{n}|\p_bh|\p_c u_{n'}\ket+
\bra \p_b u_{n}|u_{n'}\ket\bra \p_cu_{n'}|\p_ah|u_{n}\ket\right.+\nonumber\\
&+& \left.\bra u_{n}|\p_a u_{n'}\ket\bra \p_cu_{n'}|\p_bh|\p_c u_{n}\ket+\bra \p_b u_{n'}|u_{n}\ket\bra u_{n}|\p_ah|\p_c u_{n'}\ket\right).
\end{eqnarray}
A square bracket with a set of indices inside indicates an expression structurally identical to the preceding one, the \emph{only} difference being the order in which $a,b$ or $c$ appear in the derivatives. The specific order is given by the contents of the bracket.

The terms in the double sums are modified according to
$$ \bra u_{n'}|\p_a u_{n}\ket\bra u_{n}|\p_bh|\p_c u_{n'}\ket= \bra u_{n'}|\p_a u_{n}\ket\bra \p_bu_{n}|(\e_{n}-h)|\p_c u_{n'}\ket-
\bra \p_cu_{n}|\p_b\e_{n}|u_{n'}\ket\bra u_{n'}|\p_a u_{n}\ket$$
$$\bra \p_b u_{n}|u_{n'}\ket\bra \p_cu_{n'}|\p_ah|u_{n}\ket=\bra \p_bu_{n}|u_{n'}\ket\bra \p_cu_{n'}|(\e_{n}-h)|\p_a u_{n}\ket-
\bra \p_bu_{n}|u_{n'}\ket\bra u_{n'}|\p_a\e_{n}|\p_cu_{n}\ket$$
$$\bra u_{n}|\p_a u_{n'}\ket\bra \p_cu_{n'}|\p_bh|\p_c u_{n}\ket=-\bra \p_au_{n}|u_{n'}\ket\bra \p_cu_{n'}|(\e_{n}-h)|\p_b u_{n}\ket+
\bra \p_au_{n}|u_{n'}\ket\bra u_{n'}|\p_b\e_{n}|\p_cu_{n}\ket$$
$$\bra \p_b u_{n'}|u_{n}\ket\bra u_{n}|\p_ah|\p_c u_{n'}\ket=-\bra u_{n'}|\p_bu_{n}\ket\bra \p_au_{n}|(\e_{n}-h)|\p_c u_{n'}\ket+
\bra \p_cu_{n}|u_{n'}\ket\bra u_{n'}|\p_a\e_{n}|\p_bu_{n}\ket$$
These transformations make a few more summations over $n'$ possible, and we finally arrive at
\begin{eqnarray}\label{eq:me3}
Q^{inter,me}_{ab}(0, \qq)&=&e^2\sum_{\pp,n} f_{n}\frac{q_c}2(\bra\p_a u_{n}|(\p_b h+\p_b\e_{n})|\p_c u_{n}\ket+[cab]-[cba]-[bac])+\nonumber\\
&+&e^2\sum_{\pp,nn'} f_{n}\frac{q_c}2\left(\bra u_{n'}|\p_a u_{n}\ket\bra \p_bu_{n}|(\e_{n}-h)|\p_c u_{n'}\ket+\bra \p_bu_{n}|u_{n'}\ket\bra \p_cu_{n'}|(\e_{n}-h)|\p_a u_{n}\ket\right.-\nonumber\\
&-& \left.\bra \p_au_{n}|u_{n'}\ket\bra \p_cu_{n'}|(\e_{n}-h)|\p_b u_{n}\ket-\bra u_{n'}|\p_bu_{n}\ket\bra \p_au_{n}|(\e_{n}-h)|\p_c u_{n'}\ket\right)
\end{eqnarray}

\subsubsection{Total contribution of matrix elements and transition energies}

Combining Eqs.~(\ref{eq:te2}) and (\ref{eq:me3}) we obtain
\begin{eqnarray}\label{eq:Qtot}
Q_{ab}^{inter,me+te}(0, \qq)&=&e^2\sum_{n}\int(d\pp) f_{n}\frac{q_c}2(\bra\p_a u_{n}|(\p_b h+\p_b\e_{n})|\p_c u_{n}\ket+[cab]+[bca]-[acb]-[cba]-[bac])+\nonumber\\
&+&e^2\sum_{n,n'}\int(d\pp) f_{n}\frac{q_c}2\left(\bra u_{n'}|\p_a u_{n}\ket\bra \p_bu_{n}|(\e_{n}-h)|\p_c u_{n'}\ket+\bra \p_bu_{n}|u_{n'}\ket\bra \p_cu_{n'}|(\e_{n}-h)|\p_a u_{n}\ket\right.-\nonumber\\
&-& \left.\bra \p_au_{n}|u_{n'}\ket\bra \p_cu_{n'}|(\e_{n}-h)|\p_b u_{n}\ket-\bra u_{n'}|\p_bu_{n}\ket\bra \p_au_{n}|(\e_{n}-h)|\p_c u_{n'}\ket\right.-\nonumber\\
&-&\left.\bra u_{n'}|\p_a u_{n}\ket\bra\p_b u_{n} |(\e_{n'}-h)|\p_c u_{n'}\ket+\bra u_{n'}|\p_b u_{n}\ket\bra\p_a u_{n} |(\e_{n'}-h)|\p_c u_{n'}\ket\right)
\end{eqnarray}
It can be shown that the double-summation terms cancel out:
\begin{eqnarray}
\sum_{n,n'} f_{n}\frac{q_c}2&&\left(\bra u_{n'}|\p_a u_{n}\ket\bra \p_bu_{n}|(\e_{n}-h)|\p_c u_{n'}\ket+\bra \p_bu_{n}|u_{n'}\ket\bra \p_cu_{n'}|(\e_{n}-h)|\p_a u_{n}\ket\right.-\nonumber\\
&-& \left.\bra \p_au_{n}|u_{n'}\ket\bra \p_cu_{n'}|(\e_{n}-h)|\p_b u_{n}\ket-\bra u_{n'}|\p_bu_{n}\ket\bra \p_au_{n}|(\e_{n}-h)|\p_c u_{n'}\ket\right.-\nonumber\\
&-&\left.\bra u_{n'}|\p_a u_{n}\ket\bra\p_b u_{n} |(\e_{n'}-h)|\p_c u_{n'}\ket+\bra u_{n'}|\p_b u_{n}\ket\bra\p_a u_{n} |(\e_{n'}-h)|\p_c u_{n'}\ket\right)=\nonumber\\
\sum_{n} f_{n}\frac{q_c}2&&\left(
\bra \p_bu_{n}|\left(\p_c\sum_{n'}| u_{n'}\ket\bra u_{n'}|\right)(\e_{n}-h)|\p_a u_{n}\ket
-\bra \p_au_{n}|\left(\p_c\sum_{n'}| u_{n'}\ket\bra u_{n'}|\right)(\e_{n}-h)|\p_b u_{n}\ket\right)=0
\end{eqnarray}
Thus we simply get
\begin{eqnarray}
Q_{ab}^{inter,me+te}(0, \qq)&=&e^2\sum_{n}\int(d\pp) f_{n}\frac{q_c}2(\bra\p_a u_{n}|(\p_b h+\p_b\e_{n})|\p_c u_{n}\ket+[cab]+[bca]-[acb]-[cba]-[bac]).
\end{eqnarray}
Since the expression in brackets is fully antisymmetric with respect to $a$, $b$, and $c$, we can write it as
\begin{eqnarray}\label{eq:Qfinal}
Q_{ab}^{inter,me+te}(0, \qq)&=&e^2\sum_{n}\int(d\pp) f_{n}\frac{q_c}2\e_{bac}\p_\pp\cdot\bra\p_\pp u_{n}|\times(h+\e_{n})|\p_\pp u_{n}\ket.
\end{eqnarray}

\subsection{Current in the static limit}

Combining the current that corresponds to Eq.~(\ref{eq:Qfinal}) with Eq.~(\ref{eq:currentdf}) integrated by parts, we obtain
\begin{eqnarray}
j_{a}^{static}&=&e\sum_{n}\int(d\pp)f_{n}(\p_c\m_{nc})(i\qq\times{\bf A})_a-\frac{i}2 e^2\sum_{n}\int(d\pp) f_{n}\p_\pp\cdot\bra\p_\pp u_{n}|\times(h+\e_{n})|\p_\pp u_{n}\ket(i\qq\times{\bf A})_a=\nonumber\\
&=&\frac i2 e^2\sum_{n}\int(d\pp)f_{n}\p_\pp\cdot\bra\p_\pp u_{n}|\times(h-\e_{n})|\p_\pp u_{n}\ket B_a-e^2\sum_{n}\int(d\pp) f_{n}\frac{i}2\p_\pp\cdot\bra\p_\pp u_{n}|\times(h+\e_{n})|\p_\pp u_{n}\ket B_a=\nonumber\\
&=&-e^2\sum_{n}\int(d\pp)f_{n} i \p_\pp\cdot\e_{n}\bra\p_\pp u_{n}|\times|\p_\pp u_{n}\ket B_a
=-e^2\sum_{n}\int(d\pp)f_{n} \p_\pp(\e_{n}{\bf \Omega_{n}}) B_a.
\end{eqnarray}
After an integration by parts, this can be rewritten as
\begin{equation}\label{eq:staticKubo}
\jj_g^{static}=e^2\sum_{n}\int(d\pp)\e_{n}(\p_\pp f_{n}\cdot{\bf \Omega}_{n}) \BB.
\end{equation}
This result is further discussed in Appendix~\ref{sec:static}.

\subsection{Current in the dynamic limit}

In order to get the current in the dynamic limit, we just need to subtract the static contribution of the intraband response, Eq.~(\ref{eq:intracurrent}), from the total result in the static limit, since the interband part of current is independent of the orders of limit:
\begin{eqnarray}
\jj^{dynamic}_{g}&=&e^2\sum_{\pp,n}\e_{n}\left(\p_\pp f_{n}{\bf \Omega_{n}}\right) \BB+\frac{e}{  c}\sum_{\pp,n} \m_{n}\cdot(i\qq\times {\bf A})\p_{\pp}f_{n}+\frac{e}{  c}\sum_{\pp, n} (i\qq\times\m_{n})({\p_{\pp} f_{n} {\bf A}}).
\end{eqnarray}
To obtain Eq.~(\ref{eq:dynamiccurrent}), we rewrite the current as a response to the electric field $\EE=i\omega {\bf A}$:
\begin{eqnarray}
\jj^{dynamic}_g&=&\frac{e^2}{\w}\sum_{\pp,n}\e_{n}(\p_\pp f_{n}{\bf \Omega_{n}}) (\qq\times \EE)+\frac{e}{  \omega}\sum_{\pp,n} \m_{n}\cdot(\qq\times \EE)\p_{\pp}f_{n}+\frac{e}{  \omega}\sum_{\pp,n} (\qq\times\m_{n})({\p_{\pp} f_{n} \EE}),
\end{eqnarray}
which is, indeed, Eq.~(\ref{eq:gyrocurrent}).

\section{Vanishing current in the static limit}\label{sec:static}

The result of Eq.~(\ref{eq:staticKubo}) the equilibrium current in the case of a uniform $\BB$-field can be straightforwardly obtained in the context of semiclassical formalism.  Below we will simply recover known results, but nevertheless present them for completeness of our treatment. In a static magnetic field, the equilibrium distribution function at chemical potential $\mu$ is modified according to the dispersion~(\ref{eq:qpenergy}):
\begin{equation}\label{eq:f}
  f_{n\pp}(\e_{n\pp})=f_{th}(E_{n\pp});
\end{equation}
Then the equilibrium current, $\jj_{eq}$, is given by
\begin{equation}\label{eq:staticcurrent}
  \jj_{eq}=e\sum_{n}\int (d\pp) \p_\pp E_{n\pp} f_{n}(E_{n\pp})-e^2\sum_{n}\int (d\pp)(\p_\pp\e_{n\pp}\cdot{\bf \Omega}_{n\pp}) f_{n}(\e_{n\pp})\BB.
\end{equation}
We emphasize that it is the total energy $E_{n\pp}$ from Eq.(\ref{eq:qpenergy}) that enters in the first term on the right hand side of this expression, while we can use $\e_{n\pp}$ in the second term, since it is already linear in $\BB$, and we are interested in linear response.

The equilibrium current $\jj_{eq}$ vanishes in a crystal. In the dynamic limit, the first term on the right hand side of Eq.~(\ref{eq:totalcurrent}) yielded a non-zero current along the magnetic field because electron's velocity is shifted by $-\p_\pp (\m_{n\pp}\BB)$ from its unperturbed value $\p_\pp\e_{n\pp}$. However, in the static limit, this velocity shift is exactly compensated by the shift of the energy in the distribution function, Eq.~(\ref{eq:f}). Therefore, the first term vanishes identically.
The second term in Eq.~(\ref{eq:staticcurrent}) has been associated with the static limit of chiral magnetic effect in the literature~\cite{Vilenkin,Nielsen1983,Cheianov1998,Kharzeev2008,Son2013}, and it is finite for a single Weyl point. However, when the momentum integration is extended to the entire Brillouin zone, and the summation over all bands is performed, this current vanishes due to the fact that there is zero total Berry monopole charge in the Brillouin zone~\cite{Niu2013}. The simplest way to see this explicitly is to recast the expression for the chiral-magnetic-effect-related part of the static current as a Fermi surface property. Denoting the second term on the right hand side of Eq.~(\ref{eq:staticcurrent}) as $\jj_{CME}$, and integrating by parts, one obtains
\begin{equation}\label{eq:chiralcurrent}
  \jj_{CME}=e^2\sum_{n}\int (d\pp)\e_{n\pp}(\p_\pp f_{n}{\bf \Omega}_{n\pp}) \BB,
\end{equation}
where we used the fact that $\sum_{n\pp}f_{n\pp}\e_{n\pp}\p_\pp{\bf \Omega}_{n\pp}=0$, since the momentum-space divergence of the Berry curvature is non-zero only at singularities associated with band touchings, and the signs of the monopole charges are the opposite for the two degenerate bands. By switching from integration over quasimomenta to the integration over iso-energetic surfaces and energy, one obtains at zero temperature
\begin{equation}\label{eq:staticCME}
  \jj_{CME}=-\frac{e^2\mu}{4\pi^2}\BB\sum_{\text{fs}}\frac{1}{2\pi}\int d{\bf S}_{\text{fs}}\cdot{\bf \Omega}_{\text{fs}},
\end{equation}
where we switched from band summation to summation over Fermi surfaces. The orientation of $d\textbf{S}_{\text{fs}}$ has to be chosen as the outer (inner) normal for electron (hole) Fermi surfaces, because of their opposite group velocity direction. Since the signs of the Berry curvature are also opposite for electron and hole surfaces, we conclude that the contribution of a given Weyl point is independent of the position of the chemical potential relative to its nodal energy, and the current is given by
\begin{equation}\label{eq:currentcancellation}
  \jj_{CME}=\frac{e^2\mu}{4\pi^2}\BB\sum_w Q_{w}=0,
\end{equation}
where $\sum_{w}\ldots$ denotes the sum over Weyl points, and $Q_{w}=\pm 1$ is the chirality of a Weyl point.

The result of Eq.~(\ref{eq:staticCME}) is clearly of topological origin: each Berry monopole that corresponds to a Weyl point makes a contribution to the total current that depends only on its Berry charge and chemical potential, with a universal prefactor. The total static current~(\ref{eq:currentcancellation}) is also quite universal: It is a universal zero. Eqs.~(\ref{eq:staticCME}) and (\ref{eq:currentcancellation}) describe \emph{static} CME: In a static magnetic field, a Weyl point in a band structure makes a contribution to current that flows along the magnetic field, and whose magnitude is a universal quantity; however, the total current vanishes in equilibrium.

\end{widetext}
\bibliography{Weyl_references}

\begin{thebibliography}{50}
\bibitem[{not({\natexlab{b}})}]{noteonSCME}
\bibinfo{note}{The reason for the static CME current to be a part of the
  current in the dynamic limit can be traced back to the fact that this current
  appears due to interband coherence induced by the magnetic part of the
  Lorentz force, and interband coherence effects do not generally depend on the
  order of $\omega\to0$ and $q\to0$ limits. This can be seen exlpicitly by
  careful examination of the derivation based on the Kubo formula, given in the
  Appendix.}
\bibitem[{not({\natexlab{c}})}]{noteontau}
\bibinfo{note}{Strictly speaking, it would have been physically incorrect to
  substitute $\w\to\w+i/\tau$ into Eq.~(\ref{eq:lambda}). The reason is the
  presence of the first term on its right hand side, which is commonly related
  to the static CME. That term is an interband coherence effect induced by the
  magnetic part of the Lorentz force, and is of topological nature. It is
  insensitive to the value of $\w\tau$.}
\end{thebibliography}
\bibliographystyle{apsrev}
\end{document}